\documentclass[a4paper]{jpconf}
\usepackage{graphicx}

\begin{document}
\title{The role of damping for the driven  anharmonic quantum oscillator }

\author{ Lingzhen Guo$^{1,2}$, Michael Marthaler$^{1,3}$, Stephan Andr\'e$^{1,3}$ and
          Gerd Sch\"on$^{1,3}$}

\address{$^1$\ \mbox{Institut f\"ur Theoretische Festk\"orperphysik, Karlsruhe Institute of Technology, 76128} \ \\ \  \  Karlsruhe, Germany\\
$^2$\ \mbox{Department of Physics, Beijing Normal University, Beijing 100875, China}\\
$^3$\mbox{ DFG-Center for Functional Nanostructures (CFN),
Karlsruhe Institute of Technology,}\\ \ \ 76128 Karlsruhe,
Germany\\}

\ead{lzguo@tfp.uni-karlsruhe.de}

\begin{abstract}
For the model of a linearly driven quantum anharmonic oscillator,
the role of damping is investigated. We compare the position of
the stable points in phase space obtained from a classical
analysis to the result of a quantum mechanical analysis.
 The solution of the full master equation shows that the stable points
 behave qualitatively similar to the classical solution but with small modifications.
 Both the quantum effects and
 additional effects of temperature can be described by renormalizing the damping.
\end{abstract}

 In recent years, driven anharmonic oscillators have generated a large amount of interest
 because of their use as qubit readout devices
 (i.e., the Josephson bifurcation amplifier)
\cite{Siddiqi2005,Mallet2009}. All these devices operate close to
or in the quantum regime. There are two possible stable states in
phase space, which can be distinguished by their respective
amplitude and phase.  Many studies have been performed on the
 transition between the stable states \cite{Dykman1988,Risken1990,Marthaler2006,DykmanPRE2007,Ankerhold2010}
 and further effects like multiphoton resonances
\cite{Dykman2005,Peano2006} and dynamical tunneling
\cite{Wilhelm2007,Marthaler2007} make driven anharmonic
oscillators an ideal model to study thermo-dynamics and quantum
effects in a non-equilibrium system. Most bifurcation amplifiers
are operated in the limit of strong damping to speed up the
classical read-out process of the amplifier. In this paper, we
define an equivalent to the classical stable state derived from a
quantum master equation and discuss its position in phase space as
a function of damping.

\begin{figure}
\center
\includegraphics[scale=0.9]{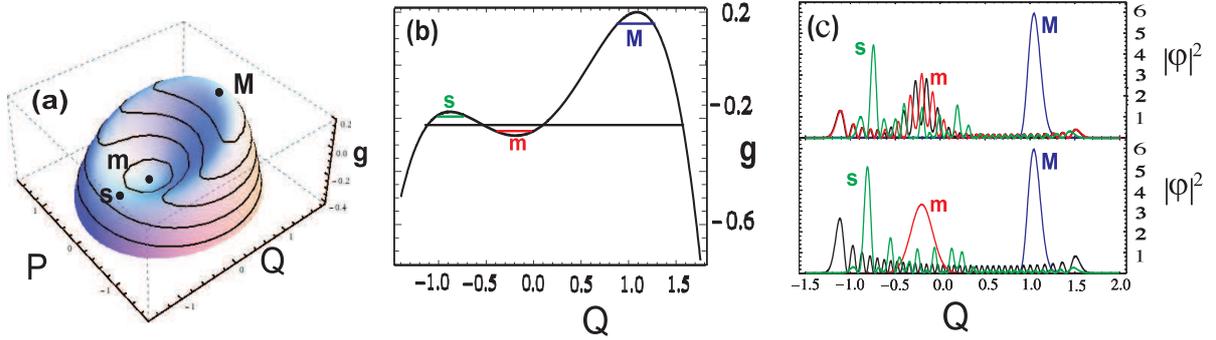}
\caption{ (a) A plot of the quasienergy $g$ in phase space. The
extrema correspond to the two stable states, i.e., the high
amplitude state (\textbf{\emph{M}}) and the low amplitude state
(\textbf{\emph{m}}). The saddle point corresponds to an unstable
state (\textbf{\emph{s}}).
 The three kinds of classical orbits are also
depicted by black closed curves. (b) A cut through the quasienergy
potential
 $g$ for $P=0$. The quantum levels nearest to maximum (\textbf{\emph{M}}),
minimum (\textbf{\emph{m}}) and saddle point (\textbf{\emph{s}})
are shown in blue, red and green short lines respectively. Shown
in long black line is the quantum level on the outer torus, which
is close to and can be generate with level \textbf{\emph{m}}. (c)
Squared amplitude of the wave functions corresponding to quantum
levels shown in Fig.1(b). We use the same colors and labels to
distinguish them. The two subfigures correspond to
 different parameters, i.e., wave function on the
outer torus is degenerate with the low amplitude state (top
subfigure) and non-degenerate (bottom subfigure). We used
$\lambda=0.027$ and $\beta=0.0341$ for this plot (degeneracy can
be created by only minor changes of $\beta$).}
\end{figure}

The driven anharmonic oscillator is described by
\begin{equation}\label{DDO}
H_S(t)=\frac{p^2}{2m}+\frac{m\Omega^2 x^2}{2} - \gamma x^4+F(t)x,
\end{equation}
where $F(t)=2F_{0}\cos(\nu t)$ stands for the external driving
force. In the rotating frame and after a rotating wave
approximation (RWA) the Hamiltonian reads
\begin{equation}\label{H}
H=\Delta a^\dagger a+\chi {a^\dagger} a({a^\dagger} a+1)+
f(a^\dagger+a).
\end{equation}
 Here, we
have introduced the detuning between the oscillator's natural
frequency and the driving frequency $\Delta=\Omega-\nu$ , $\
\chi=-3\gamma\hbar/(2m^2\Omega^3),\
f=F_0/(\hbar\Omega)\sqrt{\hbar/(2m\Omega)}$. We then introduce the
position operator $Q$ and momentum operator $P$ in the rotating
frame $ Q=\sqrt{{\lambda}/{2}}(a^\dagger + a),\ \ \
P=i\sqrt{{\lambda}/{2}}(a^\dagger - a), $ where
$\lambda=-\chi/\Delta$ is the effective Planck constant. Finally,
the scaled quasienergy Hamiltonian is defined as $
H={\Delta^2}g/\chi, $ where
\begin{equation}\label{Hq}
g=(Q^2+P^2-1)^2/4+\sqrt{\beta}Q.
\end{equation}
Here, we have neglected some constant terms. Now all the
properties of this system are dependent on one single quantity
$\beta=-2f^2\chi/\Delta^3$. The quantum properties are contained
in the commutation relation $[Q,P]=i\lambda$. The transition to
the classical regime is equivalent to the limit
 $\lambda\rightarrow 0$. The dissipative environment is taken into account using
 the Lindblad master equation
\begin{eqnarray}\label{ME}
\dot{{\rho}}(\tau)=-\frac{i}{\lambda}[H,{\rho}]
 + \eta \{ (1+\bar{n}){\cal D}[a]{\rho}
    + \bar{n}{\cal D}[a^{\dagger}]{\rho} \}    ,
\end{eqnarray}
where the Lindblad operator is defined through ${\cal
D}[A]{\rho}\equiv
A{\rho}A^{\dagger}-\frac{1}{2}\{A^{\dagger}A,{\rho} \} $,
$\bar{n}$ is the Bose-Einstein distribution, and $\eta$ is the
dimensionless damping strength \cite{DykmanPRE2007,Serban2010}.

Let's first discuss the classical dynamics of the driven Duffing
oscillator in the weak-damping limit. For $0<\beta<4/27$, the
Hamilton function exhibits three extrema in phase space,  as
indicated in Fig.1(a), i.e., a maximum $\emph{\textbf{M}}$, a
minimum $\emph{\textbf{m}}$ and a saddle point
$\emph{\textbf{s}}$. Without friction, there are three kinds of
possible periodic orbits. They correspond to different
isoenergetic sections of the Hamilton function. The orbits with an
ear-shape around the maximum $\emph{\textbf{M}}$ form one group.
Those circling the minimum $\emph{\textbf{m}}$ form another group.
The third group consists of the orbits on the outer torus. One can
calculate the explicit expressions of these orbits based on the
classical equation of motion: $\partial_t Q=\partial_P
g,\partial_t{P}=-\partial_Q g.$ In the vicinities of the extrema
$\emph{\textbf{M}}$ and
 $\emph{\textbf{m}}$ the system is equivalent to an underdamped harmonic oscillator.
  Therefore, if damping is included
 the orbits nearby
 will shrink towards the extrema.  As a result, they are stable
 points, corresponding to the oscillations with \emph{high} and \emph{low}
 amplitudes respectively. They are separated by a phase space
 barrier associated with the unstable saddle point
 $\emph{\textbf{s}}$.

 In the quantum  limit,
 the classical orbits will be
 quantized into a series of discrete energy levels. They can be calculated
 by diagonalizing the scaled Hamiltonian (\ref{Hq}), which results
 in a set of eigenvalues and eigenstates, i.e., $g|n\rangle=g_n|n\rangle$.
 In Fig.1(b) we show a cut through the quasienergy potential
 $g$ for $P=0$. The classically stable states are the extrema of the
 potential. The quantum levels corresponding to the three extrema
 are shown by colored short lines and labelled by
 \textbf{\emph{M}},\textbf{\emph{m}} and \textbf{\emph{s}}
 respectively. The black long line indicates one quantum level on
 the outer torus, which is close to and can be resonant with the level \textbf{\emph{m}}.
 Fig.1(c) shows the quantum analogues (i.e., squared amplitude of the wave
 functions, $|\varphi|^2$) of the classical orbits corresponding to the quantum levels shown in
 Fig.1(b). We use the same colors and labels in each figures.
 In principle, the small
 amplitude state is coupled via tunnelling to states on the outer
 torus. If a state on the outer and inner torus are degenerate, the eigenstates are mixed states and this results
 in a large change of the degenerate wave functions (see the red and black curves in Fig.1(c)). However, one should note that
 in most of the relevant parameter regime
 the dynamics of the system is dominated by intra-well transitions \cite{Dykman1988,Marthaler2006}.

\begin{figure}
\center
\includegraphics[scale=1]{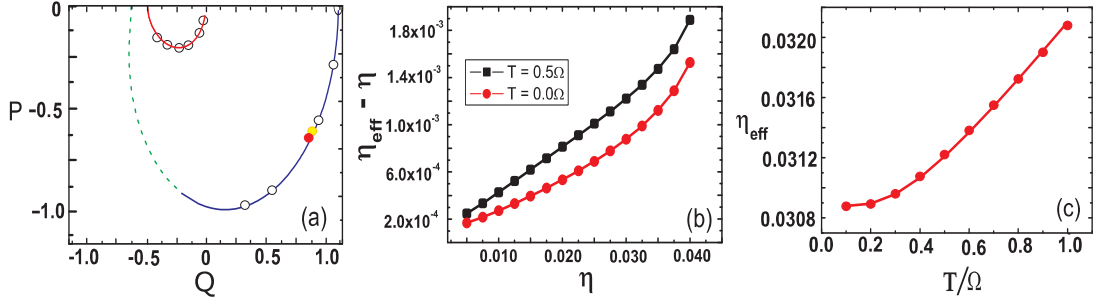}
\caption{(a) Positions of the low amplitude state (red solid
line), high amplitude state (blue solid line) and the unstable
saddle point (green dashed line) from the solution of the
classical equation of motion for finite damping and from the
solution of the master equation (circles) respectively. The black
empty circles represent zero temperature, while the yellow and red
solid circles correspond to temperatures $T=2.0\Omega$ and
$T=3.0\Omega$ respectively. (b) The difference between the
effective damping $\eta_{\rm eff}$ and real quantum simulation
damping $\eta$ at zero temperature (red solid circles) and
$T=0.5\Omega$ (black solid squares) respectively. (c) The
relationship between effective damping and temperature at fixed
damping $\eta=0.03$. Other parameters are $\lambda=0.027$ and $
\beta=0.12$. }
\end{figure}

As damping increases, the positions of the stable points in phase
space will shift. We can find the classical solution for the
stable points from the classical equation of motion including
damping:$\ \dot{Q}=\partial g/\partial P-\eta Q,\
\dot{P}=-\partial g/\partial Q-\eta P$. For finite damping the
system is classically bistable for
$\beta^{(1)}<\beta<\beta^{(2)}$, with
$\beta^{(1,2)}=2(1+9\eta^2\mp(1-3\eta^2)^{3/2})/27$
\cite{DykmanPRE2007}. In Fig.2(a), we plot the possible positions
of the low amplitude state (red solid line), high amplitude state
(blue solid line) and the unstable state (green dashed line) in
phase space.

In order to describe the damping effects in the quantum regime, we
turn to the master equation (\ref{ME}). We can also define a
stable state using our master equation. At exactly
 zero temperature, it is possible to get
 an analytical expression for the
 stationary distribution in the so called  P-representation
 \cite{Drummond1980-1,Kheruntsyan1999}, which is  one of several
 quasi-probability distributions
 \cite{Carmichael,Risken1989}. With the help of the P-representation, the
exact analytical solution for the stationary density matrix  for
$\bar{n}=0$
 can be obtained in the
photon number representation (see Eq.(15) in
Ref.\cite{Kheruntsyan1999}). We can diagonalize it and use its
eigenstates as our basis. A diagonal representation will not only
give us a clear physical picture of the dynamics of the system but
also simplifies any numerical simulation.

To understand how our solution compares to the solution of the
classical equation, we use the following procedure: (1)
diagonalize the exact solution; (2) identify the eigenstates which
correspond to the two stable states and calculate the expectation
value of position and momentum (or amplitude and phase). An
example for the results of such a calculation are the black empty
circles in Fig.2(a); (3) compare the expectation values of the
wave function with those expected from the classical equation of
motion. Since the states are located on the same curve as the
classical state we can assign each state a value $\eta_{\rm eff}$
that corresponds to the damping of the equivalent classical stable
point.

Due to the finite effective Planck constant $\lambda$ used in this
work, our results display some quantum effects. One is the
  disappearance of the small-amplitude state already for weaker damping than classically predicted.
 The reason is that near the bifurcation point, the
low amplitude well is so shallow that the zero point energy
exceeds the potential barrier. As a result, there is no quantum
level in the vicinity of the minimum {\bf m}. The situation is
similar for the large-amplitude state if $\beta$ is small.
 Another quantum signature is the renormalization of damping, $\eta_{\rm eff}\neq \eta$.
 We show a comparison of $\eta$, as chosen for the master equation
 and the effective damping $\eta_{\rm eff}$ of the equivalent classical stable point in Fig.2(b).
 As $\lambda$ becomes smaller we find $\eta_{\rm eff}\approx\eta$
\cite{DykmanPRE2007,Serban2010}.

 The effect of temperature can be included in an exact numerical solution of the master equation.
  Then we again follow the steps outlined above
 and find that the eigenstates that diagonalize the matrix have average momenta and coordinates that fall on the
 curve of the classically stable states (see yellow and red solid circles in
 Fig.2(a)).
 We find that the temperature also introduces
 a weak renormalization of the damping $\eta$. We plot the difference, $\eta_{\rm eff}-\eta$, as function of $\eta$ in Fig.2(b) at a
  finite temperature (black solid squares in
 Fig.2(b)). We also show the renormalized damping as a function of
 $T$
 for fixed $\eta$ in Fig.2(c). As one can see, there is a small quantum correction for $T=0$, and temperature
 has, as expected, no effect as long as $T\ll \Omega$. As temperature increases we get
 an additional small shift along the curve of the classically stable state. Generally we find that the states that diagonalize the
 density matrix at $T=0$ still remain an efficient diagonal basis up to $T\approx \Omega$.


\end{document}